\documentclass[aps,prl,reprint,preprintnumbers,amsmath,amssymb,floatfix]{revtex4-2}

\usepackage[T1]{fontenc}
\usepackage{lmodern}
\usepackage{microtype}
\usepackage{bm}
\usepackage{graphicx}
\graphicspath{{draft/}}
\usepackage{booktabs}
\usepackage[hidelinks]{hyperref}
\usepackage[dvipsnames]{xcolor}

\begin{document}

\title{Tests of an EP-Preserving Wake Force}
\preprint{DESY-26-149}
\preprint{LITP-26-23}

\author{Abhishek Banerjee}
\affiliation{Deutsches Elektronen-Synchrotron DESY,
Notkestr. 85, 22607 Hamburg, Germany}
\affiliation{Maryland Center for Fundamental Physics, Department of Physics, University of Maryland, College Park, MD 20742, USA}

\author{Dawid Brzeminski}
\affiliation{Leinweber Institute for Theoretical Physics,
Department of Physics, University of Michigan,
Ann Arbor, MI 48109, USA}
\date{\today}

\begin{abstract}
A universal quadratic coupling of ultralight scalar dark matter to ordinary
matter preserves the weak equivalence principle for unscreened point
particles. A finite test body is nevertheless not a passive probe: it distorts the scalar field that accelerates it. 
We show that this field distortion produces a differential acceleration between concentric test masses, even when their microscopic coupling is universal.
We calculate this signal including how each test mass distorts the field acting on the other, deriving the weak-screening limit and extending the calculation to stronger screening using concentric spherical layers. Applying the model to MICROSCOPE, we find differential signals in both its equivalence-principle sensor SUEP, whose test masses have different compositions, and its reference sensor SUREF, whose test masses have the same composition. We combine both sensors with the dark matter wake near Earth to recast the published MICROSCOPE measurements. Within this model, existing measurements probe scalar masses up to approximately $8 \times 10^{-9}\,$eV and couplings below the supernova bound. An analysis using the additional frequency components of the scalar signal could extend the mass reach and probe couplings below the BBN bound.

\end{abstract}

\maketitle

\section{Introduction}

A universally coupled force with an inverse-square dependence is difficult to distinguish from gravity~\cite{PhysRev.126.1875,Kaplan:2000hh,Sundrum:2003yt}. A standard search strategy is to exploit its finite range, looking for departures from the inverse-square law~\cite{Fischbach:1996eq,Adelberger:2003zx,Lee:2020zjt}. Light dilatons provide a well-studied example of scalar-mediated
forces with both composition-independent and composition-dependent
contributions~\cite{Fischbach:1996eq,Kaplan:2000hh,Damour:2010rp}.
Forces generated by scalar dark matter with universal quadratic couplings to matter face a similar challenge. Earth distorts the ambient dark matter field, producing a wake force whose spatial dependence can differ from gravity~\cite{Hees:2018fpg,Banerjee:2022sqg,VanTilburg:2024xib, Barbosa:2024pkl, Gan:2025nlu,Brzeminski:2026pgz,Bouley:2026frx}. Quadratic couplings also allow matter to screen the field~\cite{Olive:2007aj,Hees:2018fpg,Banerjee:2025dlo,Gan:2025nlu}. 
For positive couplings, this screening can suppress the force near Earth’s surface, motivating experiments in space. However, even there the satellite housing can suppress the signal reaching the apparatus~\cite{Burrage:2025grx}.

Screening also offers a way to distinguish a universal dark matter force from gravity. Bodies with different internal structures can distort the field differently and therefore acquire different accelerations, even when their microscopic coupling is identical. 
Previous work proposed exploiting this effect by comparing the motion of two satellites with different internal structures~\cite{Burrage:2025grx}. 
Here we show that separate satellites are not required. Concentric test masses can also develop a differential signal. Each mass modifies the field acting on both, so their accelerations must be calculated within a common field solution. This makes experiments such as MICROSCOPE~\cite{MICROSCOPE:2022doy,Touboul:2022yrw} sensitive to a wake force, even for a universally coupled scalar.

MICROSCOPE was designed to search for violations of the weak equivalence principle by comparing the accelerations of concentric test masses. It carried two pairs, SUEP with masses of different compositions and SUREF with masses of the same composition~\cite{Liorzou:2020vob}. If the masses were passive probes of the same field, a universal coupling would produce no differential acceleration in either pair. Even weak scattering changes this expectation. The outer mass modifies the field gradient experienced by the inner mass, allowing their accelerations to differ. This effect depends on the dimensions and densities of the masses, without requiring composition-dependent microscopic couplings. Both SUEP and SUREF can therefore be sensitive to a wake force from universal couplings. 

Interference between dark matter waves scattered by MICROSCOPE's nested test masses has also been studied for momentum transfer forces at much higher dark matter masses~\cite{Fu:2026atp}. 
Related studies in chameleon theories have examined how nested
test masses and surrounding material shape the scalar
field~\cite{Pernot-Borras:2019gqs,Pernot-Borras:2020jev}.
MICROSCOPE calibration measurements have also been used to
constrain the resulting displacement-dependent
force~\cite{Pernot-Borras:2021edr}.
In the regime considered here, the dark matter wavelength is much larger than the apparatus, and the differential acceleration
arises from the combined distortion of the dark matter field by
the Earth and the nested test masses.

In this {\it letter}, we derive the differential acceleration of concentric test masses, including their mutual screening. Within a spherical-shell approximation, we obtain the leading weak-screening contribution and extend the calculation to arbitrary screening strength. We use this description to assess the joint sensitivity of SUEP and SUREF to the Earth-induced wake force, and revisit the complementary sensitivity from LAGEOS II orbital precession~\cite{Burrage:2026loe}.

\section{Screening in the long-wavelength regime}
\label{sec:screening}
We begin by introducing the universal quadratic interaction that governs both the force on the test masses and the propagation of the dark matter field.
Quadratic scalar interactions are often organized by their couplings to
individual Standard Model (SM) operators. We instead focus on a universal
interaction, for which every matter sector couples to the same conformally
rescaled metric. With metric signature $(+,-,-,-)$, we write
\begin{equation}
 \begin{split}
 S={}&\int d^4x\sqrt{-g}\left[
 \frac{1}{2}g^{\mu\nu}\partial_\mu\phi\,\partial_\nu\phi
 -\frac{1}{2}m_{\rm DM}^2\phi^2\right] \\
 &+S_m\!\left[A^2(\phi)g_{\mu\nu},\Psi_m\right],
 \end{split}
 \label{eq:universal_action}
\end{equation}
where the conformal factor is
\begin{equation}
 A(\phi)=1+\frac{\kappa^2d}{2}\phi^2
 +\mathcal O(\kappa^4\phi^4)\,.
 \label{eq:conformal_factor}
\end{equation}
Here $\kappa^2=4\pi G_N$ and $d$ is a dimensionless
coupling. 

The universal coupling makes the mass of a nonrelativistic body depend on the local scalar field,
\begin{equation}
 m_A(\phi)=m_{A,0}A(\phi)
 =m_{A,0}\left[1+\frac{\kappa^2d}{2}\phi^2
 +\mathcal O(\kappa^4\phi^4)\right],
 \label{eq:universal_mass}
\end{equation}
where $m_{A,0}$ is the mass at $\phi=0$. A spatial variation of the scalar field therefore produces a force. For a point-like unscreened body, the oscillation-averaged acceleration is
\begin{equation}
 \mathbf a_\phi=-\frac{\kappa^2d}{2}
 \boldsymbol\nabla\langle\phi^2\rangle \,.
 \label{eq:universal_acceleration}
\end{equation}
This acceleration is independent of composition, so unscreened point-like bodies obey the weak equivalence principle. 

Ordinary matter also changes the propagation of the scalar. In a
nonrelativistic medium of energy density $\rho_{\rm SM}(\mathbf x)$, the
Klein--Gordon equation is
\begin{equation}
 \left[\Box+m_{\rm DM}^2(\mathbf x)\right]\phi=0,
 \qquad
 m_{\rm DM}^2(\mathbf x)=m_{\rm DM}^2
 +\Delta m_{\rm DM}^2(\mathbf x),
 \label{eq:matter_kg}
\end{equation}
with
\begin{equation}
 \Delta m_{\rm DM}^2(\mathbf x)
 =d\,\kappa^2\rho_{\rm SM}(\mathbf x)\,.
 \label{eq:matter_mass_shift}
\end{equation}
We focus on positive couplings, for which matter increases the effective scalar mass and acts as a barrier to propagation.

Earth distorts the surrounding dark matter field on scales set by its radius and the dark matter wavelength. We consider an experiment with characteristic size $R_{\rm exp}$ satisfying
\begin{equation}
 R_{\rm exp}\ll R_\oplus,\lambda_{\rm DM},
 \label{eq:experimental_scale_hierarchy}
\end{equation}
without assuming a hierarchy between $R_\oplus$ and
$\lambda_{\rm DM}$. Consequently, even when the global field profile is
nontrivial, it can be expanded locally across the experiment as a constant
plus a linear gradient, with higher spatial derivatives neglected. We
organize this local field in partial waves. Since
$2\pi R_{\rm exp}/\lambda_{\rm DM}\ll1$, the $\ell=0$ (monopole) and $\ell=1$ (dipole) modes capture the
leading constant and gradient components.

To identify the leading differential signal, we first consider the weak-screening limit,
\begin{equation}
 \left|\Delta m_{\rm DM}^2\right|R_{\rm exp}^2\ll 1\,,
 \label{eq:weak_screening_condition}
\end{equation}
in which the field distortion generated by the experimental apparatus can
be treated perturbatively. This condition is independent of the long-wavelength approximation. We subsequently extend the calculation to arbitrary screening strength.

We decompose the Earth-distorted dark matter field into monochromatic
components. For each component, we write
$\tilde\phi(\mathbf x,t;\mathbf k)
=\operatorname{Re}[e^{-i\omega t}\psi(\mathbf x;\mathbf k)]$,
with $\omega^2=m_{\rm DM}^2+k^2$. We expand its spatial profile into
multipoles about the center of the experiment. Retaining the leading
monopole and dipole terms, we write, for the spherical model considered
below,
\begin{equation}
 \psi(\mathbf x;\mathbf k)
 =\psi_0(\mathbf k)u_0(r)
 +\psi_1(\mathbf k)u_1(r)\cos\theta+\cdots .
 \label{eq:local_field_modes}
\end{equation}
Here $r$ is measured from the center of the experiment, and $\theta$ is measured from the local $z$ axis, chosen along the direction of the linearly varying component. The coefficients $\psi_0$ and $\psi_1$
specify the constant and linearly varying parts of the field incident
on the apparatus. The corresponding radial profiles $u_0$ and $u_1$
account for propagation through the apparatus. We normalize both
profiles to be dimensionless, with unscreened long-wavelength limits
$u_0=1$ and $u_1=r/R_{\rm exp}$, so that $\psi_0$ and $\psi_1$ have
the same dimensions.

At leading order in the long-wavelength expansion, these radial
profiles are independent of momentum. The averaging over incoming
modes therefore acts only on the incident coefficients. After
averaging over oscillations and independent phases, their quadratic
combinations determine the local incident intensity and gradient:
\begin{equation}
 \begin{aligned}
 \left.\langle\phi^2\rangle_{\rm ext}\right|_{\mathbf x=0}
 &=\frac12\langle|\psi_0|^2\rangle_{\mathbf k}\,,\\
 \left[\partial_z\langle\phi^2\rangle\right]_{\rm ext}
 &=\frac{\operatorname{Re}
 \langle\psi_0^*\psi_1\rangle_{\mathbf k}}{R_{\rm exp}}\,.
 \end{aligned}
 \label{eq:external_phi_squared_gradient}
\end{equation}
Here $\langle\cdots\rangle_{\mathbf k}$ denotes an average weighted
by the incoming momentum distribution. The subscript ``ext''
refers to the Earth-distorted field before it interacts with the
apparatus. Its incident intensity and gradient can thus be treated
separately from the screening by the experimental masses.

\section{Concentric test masses and signal scaling}
\label{sec:concentric_masses}

We now calculate the differential acceleration of two concentric
test masses, accounting for the field distortion caused by both
masses and the surrounding apparatus. As an analytic example,
we model the apparatus as concentric homogeneous spherical layers.
Layer $i$ occupies $R_{i,-}<r<R_{i,+}$ and has density $\rho_i$.
Layers $j=1,2$ represent the test masses, while additional layers
describe surrounding material. All layers enter the common field
solution, but only the accelerations of the two test masses enter
the differential observable.

\begin{figure}[t]
 \includegraphics[width=\columnwidth]
 {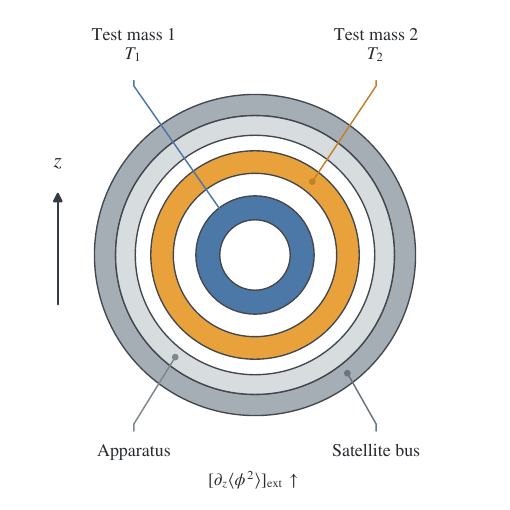}
  \caption{Schematic of the four-layer spherical model. The blue and orange shells represent the inner and outer test masses, while the two gray layers represent the surrounding apparatus and satellite bus. All four layers modify the common scalar field, but only the accelerations of the two test masses enter the differential observable. Each test mass samples the volume-averaged field gradient through its screening factor $T_j$, defined in Eq.~\eqref{eq:shell_factor_definition}. The illustrated radii and thicknesses are schematic.}
 \label{fig:concentric_shell_gradient_schematic}
\end{figure}

The acceleration of test mass $j$ depends on the volume-averaged
gradient of $\langle\phi^2\rangle$ within its material. We compare
this average with the incident gradient by defining the
dimensionless factor
\begin{equation}
 T_j(d)\equiv
 \frac{\left\langle
 \partial_z\langle\phi^2\rangle
 \right\rangle_j}
 {\left[\partial_z\langle\phi^2\rangle\right]_{\rm ext}}\,,
 \label{eq:shell_factor_definition}
\end{equation}
where the outer brackets denote a volume average over test mass $j$. 
Fig.~\ref{fig:concentric_shell_gradient_schematic} illustrates a four-layer realization of this model, with two test-mass shells surrounded by apparatus and satellite-bus layers. We use this arrangement for MICROSCOPE, with the parameters specified in App.~\ref{app:multilayer}.

Using Eq.~\eqref{eq:universal_acceleration}, the differential
acceleration of the two test masses is
\begin{equation}
 \begin{split}
 \Delta a={}&-\frac{\kappa^2d}{2}
 \left[\partial_z\langle\phi^2\rangle\right]_{\rm ext}
 \Delta T(d)\,,\\
 \Delta T(d)={}&T_1(d)-T_2(d)\,.
 \end{split}
 \label{eq:differential_acceleration}
\end{equation}
Differences in screening between the two test masses can therefore
produce a relative acceleration even when the microscopic coupling
is universal.

Without screening, both test masses have $T_j(0)=1$, so $\Delta T(0)=0$. Therefore, the weak-screening expansion begins at
first order in $d$:
\begin{equation}
 \Delta T(d)=d\left.
 \frac{\partial\Delta T}{\partial d}\right|_{d=0}
 +\mathcal O(d^2)\,.
 \label{eq:weak_shell_scaling}
\end{equation}
At fixed incident gradient, the differential acceleration is therefore
\begin{equation}
 \Delta a=-\frac{\kappa^2}{2}
 \left[\partial_z\langle\phi^2\rangle\right]_{\rm ext}
 \left.\frac{\partial\Delta T}{\partial d}\right|_{d=0}
 d^2+\mathcal O(d^3)\,.
 \label{eq:weak_differential_scaling}
\end{equation}
One power of $d$ comes from the universal force, while the other
comes from the field distortion that gives the two test masses
different screening factors.

To illustrate this scaling, we consider two concentric spherical
test masses with densities $\rho_1$ and $\rho_2$ and radii
$R_{1,-}<R_{1,+}<R_{2,-}<R_{2,+}$. The region inside the inner
shell, the gap between the shells, and the exterior are vacuum.
We write $R_{j,\pm}=R_j\pm\delta R_j/2$, where $R_j$ and
$\delta R_j$ are the mean radius and thickness of shell $j$.
In the thin-shell limit $\delta R_j/R_j\ll1$, the masses are
$M_j\simeq4\pi\rho_jR_j^2\delta R_j$. Using the first-order
field profiles derived in App.~\ref{app:weak_shells},
we obtain
\begin{equation}
 \begin{split}
 \left.\frac{\partial\Delta T}{\partial d}\right|_{d=0}
 \simeq{}&-\kappa^2\left[\rho_1R_1\delta R_1
 \left(1-\frac{2R_1}{3R_2}\right)
 +\frac{\rho_2R_2\delta R_2}{3}\right]\\
 ={}&-G_N\left[\frac{M_1}{R_1}
 \left(1-\frac{2R_1}{3R_2}\right)
 +\frac{M_2}{3R_2}\right].
 \end{split}
 \label{eq:two_shell_weak_coefficient}
\end{equation}

The calculation extends to arbitrary screening strength and
additional concentric layers, using the matching construction
given in App.~\ref{app:multilayer}. At a sufficiently strong coupling,
the individual screening factors are strongly suppressed,
and $\Delta T$ is no longer linear in $d$.

We now combine the screening factors with the gradient of
the Earth-distorted field in
Eq.~\eqref{eq:differential_acceleration} to calculate the
differential signal in MICROSCOPE.

\section{MICROSCOPE}
\label{sec:microscope}

MICROSCOPE carried two differential accelerometers, SUEP and SUREF,
each containing a pair of concentric cylindrical test masses.
SUEP contains an inner PtRh10 test mass and an outer TA6V test mass,
while both SUREF test masses are made of PtRh10.
SUEP was designed to test the equivalence principle by comparing
different compositions, with SUREF serving as a reference
sensor~\cite{Liorzou:2020vob}.

Electrostatic forces hold the test masses at fixed positions
relative to the instrument. The control forces are used to infer
the differential acceleration along the common cylinder axis.
For the scalar force considered here, scattering by the finite
test masses can produce a differential acceleration in both
sensors, including SUREF despite its identical compositions.

We calculate the differential acceleration using
Eq.~\eqref{eq:differential_acceleration}, with separate screening
factors $\Delta T_{\rm SUEP}(d)$ and $\Delta T_{\rm SUREF}(d)$.
The external gradient $[\partial_z\langle\phi^2\rangle]_{\rm ext}$
includes scattering of the dark matter field by the Earth.
Appendix~\ref{app:multilayer} describes the apparatus model,
while Appendix~\ref{app:MICROSCOPE} gives the Earth-field
calculation and its projection onto the sensitive axis.

MICROSCOPE measures the differential acceleration along an axis fixed in the satellite, so orbital motion and satellite rotation modulate the projection of the force onto this axis.
For $m_{\rm DM}\lesssim10^{-11}\,$eV, the force is
approximately radial. Orbital motion produces a modulation
at $f_{\rm orb}$, while the satellite's rotation shifts the
measured signal to $f_{\rm EP}=f_{\rm orb}+f_{\rm spin}$,
which we refer to as the EP band.
Here we use the results of MICROSCOPE's analysis that searched
for a radial force, so our recast uses the component of the
scalar signal in this band.
At higher masses, the field distorted by the Earth develops significant
angular structure and the force is no longer purely radial.
The signal is then redistributed among frequency bands at
$f=qf_{\rm orb}+f_{\rm spin}$, with the EP-band corresponding to $q=1$ and the leading sidebands
corresponding to $q=0$ and $q=2$, see Refs.~\cite{Brzeminski:2026pgz,Bouley:2026frx}.

We jointly analyze the 19 SUEP and 13 SUREF in-phase EP coefficients
obtained with ADAM and reported in Tables 7 and 6 of
Ref.~\cite{Touboul:2022yrw}. We denote these measurements by
$\widehat\eta_{c,i}$ and their statistical uncertainties by
$\sigma_{c,i}$, where $c$ labels the sensor and $i$ the data segment.
The predicted EP coefficient is
\begin{equation}
 \eta_{c,i}^{(1)}
 =\frac{\rho_{\rm DM}\kappa^2}
 {2m_{\rm DM}^2 g(r_{\rm sat})R_\oplus}
 d\,\Delta T_c(d)\,\langle A_1(t)\rangle_i\,,
 \label{eq:microscope_segment_mean}
\end{equation}
where $g(r_{\rm sat})$ is the gravitational acceleration at the
satellite and $\langle A_1(t)\rangle_i$ denotes the average over
segment $i$ of the EP-band coefficient defined in
Eq.~\eqref{eq:appendix_ep_band_amplitude}.
The projection and segment averaging follow
Ref.~\cite{Brzeminski:2026pgz} and are described in
App.~\ref{app:MICROSCOPE}.

We assume that each data segment samples the ensemble-averaged
signal, neglecting fluctuations between field realizations.
We treat the quoted statistical uncertainties as independent
Gaussian errors and use them to determine the fit weights. Holding the predicted signal template fixed at each
tested $(m_{\rm DM},d)$, the weighted fit gives
\begin{align}
 \widehat d(m_{\rm DM},d)
 &=d\,\frac{\sum_{c,i}\eta_{c,i}^{(1)}\widehat\eta_{c,i}/\sigma_{c,i}^2}
 {\sum_{c,i}(\eta_{c,i}^{(1)}/\sigma_{c,i})^2}\,,\nonumber\\
 \sigma_d(m_{\rm DM},d)
 &=\frac{d}{\sqrt{\sum_{c,i}(\eta_{c,i}^{(1)}/\sigma_{c,i})^2}}\,.
 \label{eq:microscope_joint_fit}
\end{align}

We also account for the systematic uncertainties reported for
each data segment. Their contribution $S_d$ to the uncertainty
on the coupling is calculated in
Appendix~\ref{app:MICROSCOPE}.
Combining it with the statistical uncertainty, we use
\begin{equation}
 u_d=\sqrt{\sigma_d^2+S_d^2},
 \qquad d>|\widehat d|+2u_d,
 \label{eq:microscope_budget_exclusion}
\end{equation}
to determine the excluded coupling range.

\begin{figure}[t]
 \includegraphics[width=\columnwidth]
 {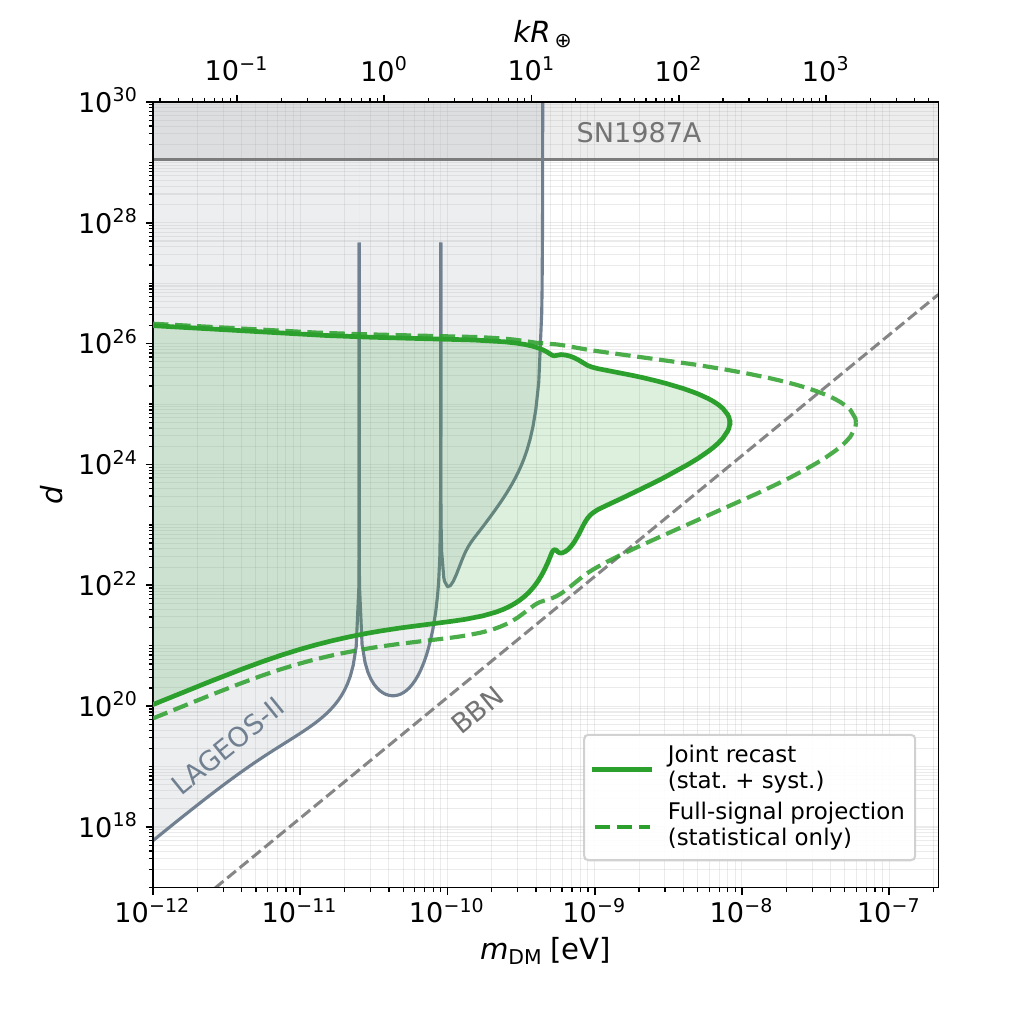}
\caption{Sensitivity to the universal coupling $d$ as a function of the scalar mass
$m_{\rm DM}$, assuming the scalar accounts for all dark matter.
The solid green curve gives the joint SUEP--SUREF recast,
including statistical uncertainties and propagated systematic
budgets.
The dashed green curve shows the projected sensitivity using the full
scalar signal and statistical uncertainties only.
Both include apparatus screening in the four-layer spherical model.
The blue-gray curve shows the revisited LAGEOS II
sensitivity~\cite{Burrage:2026loe}.
The remaining gray curves represent SN1987A (solid) and BBN (dashed) bounds
(see {\it e.g.}~\cite{Stadnik:2015kia,Sibiryakov:2020eir,Raffelt:1990yz,Olive:2007aj} and Refs. therein for more details).}
 \label{fig:microscope_universal_bound}
\end{figure}

\section{Discussion}

For a composition-dependent EP-violating force, the identical
compositions of the SUREF test masses make their differential
acceleration a reference channel.
This cancellation does not apply to the geometry-dependent
scalar force considered here, which can produce a signal
in both sensors.
We therefore compare the scalar amplitudes inferred separately
from SUEP and SUREF, accounting for their screening factors
and the geometry sampled by each segment.
Including statistical and systematic uncertainties, the
amplitudes differ by less than twice their combined uncertainty
throughout the sampled parameters.
App.~\ref{app:MICROSCOPE} describes this comparison.

Fig.~\ref{fig:microscope_universal_bound} shows the joint
SUEP--SUREF recast for $m_{\rm DM}\geq10^{-12}\,$eV and the
projected sensitivity obtained by including the sidebands.
The recast probes scalar masses
up to $m_{\rm DM}\simeq1\times10^{-8}\,$eV, while including the sidebands potentially extends the sensitivity to
$m_{\rm DM}\simeq7\times10^{-8}\,$eV. 
For the projection, we assume zero central values and independent
errors, assigning each frequency band the statistical EP-band
uncertainty of the same data segment.
Both calculations use the same field and apparatus models.
The full signal projection is described in
Appendix~\ref{app:MICROSCOPE}.

Screening raises the coupling required for detection and suppresses the signal again at strong coupling, producing a finite interval of sensitivity. As the dark matter mass increases, the unscreened sensitivity reaches couplings where screening becomes strong. Eventually, the acceleration inside the satellite remains below the experimental sensitivity for every coupling, closing the interval. Including the sidebands captures additional signal power outside the EP band and can extend the mass reach before screening again prevents detection.

For comparison, Fig.~\ref{fig:microscope_universal_bound} also
shows the sensitivity inferred from LAGEOS II pericenter
precession~\cite{Burrage:2026loe}. We calculate the induced precession including the
angular structure of the field near Earth, satellite
screening, and the changing orbital orientation.
Details of the calculation are outlined in App.~\ref{app:lageos}.
The MICROSCOPE sensitivity extends to higher masses than the
LAGEOS II comparison and probes couplings well below the
SN1987A boundary~\footnote{We recast the SN1987A bound from Ref.~\cite{Sibiryakov:2020eir}, which was derived using the results of Refs.~\cite{Raffelt:1990yz,Olive:2007aj,Hanhart:2000er}. We note that the underlying cooling argument was critically examined in Ref.~\cite{Bar:2019ifz}}.  
MICROSCOPE provides a complementary test using the local dark matter field, while the stronger BBN bound relies on cosmological assumptions~\footnote{
BBN bounds derived in Refs.~\cite{
Stadnik:2015kia, Sibiryakov:2020eir, Bouley:2022eer} relate the present-day scalar abundance to the field evolution
during nucleosynthesis within a specified cosmological model.
These constraints therefore depend on the assumed scalar
potential, interactions, and early-Universe history.
They need not apply unchanged if the scalar sector emerges only
after BBN~\cite{compscalar:2026}, or if departures from quadratic
interactions limit changes in particle masses, as illustrated
for variations of the electron mass in Ref.~\cite{Delaunay:2025pho}.}. 
Our projection indicates that an analysis including the additional sidebands should improve on the BBN bound at higher masses.

The masses, compositions, and dimensions of the test bodies
are known, together with the total mass and dimensions of the
experimental apparatus and an approximate satellite mass.
The available description does not specify the full surrounding
mass distribution.
Our concentric spherical model incorporates the known
concentration of material near the test masses, while spreading
the remaining mass over the satellite volume.
This model is neither necessarily conservative nor optimistic.
If more of the remaining mass lies close to the experiment,
screening can suppress the signal more strongly than shown in
Fig.~\ref{fig:microscope_universal_bound}.
App.~\ref{app:multilayer} describes the adopted layers
and their parameters.

Spherical symmetry also excludes internal gradients generated
by an asymmetric mass distribution.
At leading order in the local long-wavelength expansion,
the contribution generated from the spatially uniform incident
field is fixed relative to the satellite.
Its variation follows the incident field intensity along the
orbit, without the spin modulation of the Earth-induced gradient.
It therefore contributes at orbital harmonics rather than
the EP band used in our recast.
The spherical model is intended to approximate the screening
of the external gradient.

These results show how experiments designed to test the equivalence principle can
also probe universally coupled scalar dark matter through
the mutual screening of their test masses.
For concentric bodies, the differential signal depends on
their dimensions and densities even when their microscopic
couplings are identical.
The effect is strong enough that even the existing analysis,
which uses only part of the signal, provides constraints
competitive with other experimental probes.
\section*{Acknowledgments}
We thank Aaron Pierce for comments on the draft.
The research of AB is supported partly by the Deutsche Forschungsgemeinschaft under Germany’s Excellence Strategy EXC 2121 “Quantum Universe” – 390833306, as well as from
the National Science Foundation under grant number
PHY-2514660 and the Maryland Center for Fundamental Physics.
D.B. is supported by the Department of Energy under grant number
DE-SC0007859 and in part by a Leinweber postdoctoral fellowship.
The authors gratefully acknowledge the support of the Aspen Center for Physics, which is
supported by the National Science Foundation grant PHY-2210452, where this work was initiated.

\setcounter{secnumdepth}{1}
\appendix
\section{Weak-screening shell profiles}
\label{app:weak_shells}
\setcounter{equation}{0}

In this section, we derive the weak-screening coefficient
$\left.\partial_d\Delta T\right|_{d=0}$ for two concentric
spherical test masses. In the long-wavelength limit
$k_{\rm DM}R_{\rm exp}\ll1$, we expand the radial field profiles
to first order in $d$
\begin{equation}
 \begin{aligned}
  u_0(r)&=1+d\,u_0^{(1)}(r)+\mathcal O(d^2)\,,\\
  u_1(r)&=\frac{r}{R_{\rm exp}}
  +d\,u_1^{(1)}(r)+\mathcal O(d^2)\,.
 \end{aligned}
 \label{eq:weak_mode_expansion}
\end{equation}
The zeroth-order profiles $(r/R_{\rm exp})^\ell$ are the
known unscreened solutions for $\ell=0,1$. Substituting this
ansatz into the radial field equation and retaining terms
linear in $d$ gives
\begin{equation}
 \frac{d^2u_\ell^{(1)}}{dr^2}
 +\frac{2}{r}\frac{du_\ell^{(1)}}{dr}
 -\frac{\ell(\ell+1)}{r^2}u_\ell^{(1)}
 =\kappa^2\rho(r)\left(\frac{r}{R_{\rm exp}}\right)^\ell,
 \label{eq:weak_profile_equation}
\end{equation}
where $\rho(r)$ is the density profile of the experimental
apparatus. The source on the right-hand side is therefore
determined by the known unscreened profile. Requiring the field to be smooth at the origin, matching the field and its radial derivative across material interfaces, and holding the incident coefficients $\psi_0$ and $\psi_1$ fixed gives
\begin{equation}
 \begin{split}
 u_\ell^{(1)}(r)
 =-\frac{1}{(2\ell+1)R_{\rm exp}^{\ell}}
 \bigg[&r^{-\ell-1}
 \int_0^r ds\,\kappa^2\rho(s)s^{2\ell+2}\\
 &+r^\ell\int_r^\infty ds\,\kappa^2\rho(s)s\bigg].
 \end{split}
 \label{eq:weak_green_solution}
\end{equation}
The two integrals account for material inside and outside
radius $r$, respectively.

The leading contribution to $\partial_z\langle\phi^2\rangle$
is proportional to $\partial_z[u_0(r)u_1(r)\cos\theta]$. For a spherical
test mass, the divergence theorem reduces the volume average
of this contribution to surface terms. Defining $w(r)=u_0(r)u_1(r)$,
we obtain
\begin{equation}
 T_j(d)=R_{\rm exp}
 \frac{R_{j,+}^2w(R_{j,+})-R_{j,-}^2w(R_{j,-})}
 {R_{j,+}^3-R_{j,-}^3}\,.
 \label{eq:shell_factor}
\end{equation}
Without screening, $w(r)=r/R_{\rm exp}$ and $T_j=1$ for
both test masses. Differentiating Eq.~\eqref{eq:shell_factor} at $d=0$ gives the screening correction directly in terms
of the first-order field profiles
\begin{equation}
 \begin{split}
 \left.\frac{\partial T_j}{\partial d}\right|_{d=0}
 ={}&
 \frac{R_{j,+}^3u_0^{(1)}(R_{j,+})
       -R_{j,-}^3u_0^{(1)}(R_{j,-})}
      {R_{j,+}^3-R_{j,-}^3}\\
 &+R_{\rm exp}
 \frac{R_{j,+}^2u_1^{(1)}(R_{j,+})
       -R_{j,-}^2u_1^{(1)}(R_{j,-})}
      {R_{j,+}^3-R_{j,-}^3}\,.
 \end{split}
 \label{eq:weak_shell_boundary}
\end{equation}
Only the values of these profiles at the two boundaries of
test mass $j$ enter its screening factor.

For the two concentric test masses, we take $\rho(r)=\rho_j$
within $R_{j,-}<r<R_{j,+}$, for $j=1,2$, and zero elsewhere.
We write $R_{j,\pm}=R_j\pm\delta R_j/2$.
Substituting Eq.~\eqref{eq:weak_green_solution} into
Eq.~\eqref{eq:weak_shell_boundary} and expanding in
$\delta R_j/R_j\ll1$ gives
\begin{equation}
 \begin{aligned}
 \left.\frac{\partial T_1}{\partial d}\right|_{d=0}
 &\simeq-\kappa^2\left[
 \rho_1R_1\delta R_1+\frac43\rho_2R_2\delta R_2
 \right],\\
 \left.\frac{\partial T_2}{\partial d}\right|_{d=0}
 &\simeq-\kappa^2\left[
 \frac{2\rho_1R_1^2\delta R_1}{3R_2}
 +\rho_2R_2\delta R_2
 \right].
 \end{aligned}
 \label{eq:two_shell_individual_factors}
\end{equation}
Taking the
difference, with $\Delta T=T_1-T_2$, gives
\begin{equation}
 \begin{split}
 \left.\frac{\partial\Delta T}{\partial d}\right|_{d=0}
 \simeq-\kappa^2\bigg[&\rho_1R_1\delta R_1
 \left(1-\frac{2R_1}{3R_2}\right)\\
 &+\frac{\rho_2R_2\delta R_2}{3}\bigg].
 \end{split}
 \label{eq:appendix_two_shell_reduction}
\end{equation}

\section{Screening model}
\label{app:multilayer}
\setcounter{equation}{0}

The publicly available information is insufficient for a
detailed reconstruction of the MICROSCOPE mass distribution.
We therefore construct a simplified three-dimensional model to assess whether concentric spherical layers capture the screening relevant for the sensitivity calculation.
The wavelength of dark matter is much larger than the satellite, so the externally supplied field varies slowly across the
experiment.
Its propagation through the apparatus nevertheless depends
on the distribution of matter, particularly the dense material
near the test masses.

The three-dimensional model comprises a cuboid satellite bus
and two side-by-side sensor assemblies, each containing
concentric cylindrical test masses and surrounding hardware.
This model incorporates the known test-mass geometry and
the concentration of material around the sensors.
We calculate the screening in SUEP and SUREF using finite
elements and compare it with concentric spherical models.

\subsection{Apparatus geometry}

We retain the hollow cylindrical shapes of the four test
masses, using the dimensions and material densities reported
in Ref.~\cite{Liorzou:2020vob}.
SUEP contains an inner PtRh10 mass and an outer TA6V mass,
while both SUREF masses are PtRh10.
We take the common cylinder axis to be the $z$ axis and
place the two sensor centers symmetrically about the origin,
with an assumed separation of $18\,$cm.

We represent the hardware surrounding each sensor by a
homogeneous cylindrical region with a central vacuum region
containing the two test masses.
We take the effective outer radius and length to be $8.8\,$cm
and $34.8\,$cm, respectively.
This construction concentrates the surrounding material
near the test masses while averaging over the individual
electrodes, supports, and enclosing components.
We assign the combined sensor assembly a mass of
$25\,$kg~\cite{Liorzou:2020vob}, subtract the masses of the four modeled test
bodies, and divide the remaining mass equally between the
two surrounding hardware regions.
Their densities follow from the assigned masses and
material volumes.

\begin{figure}[t]
 \centering
 \includegraphics[width=\columnwidth]
 {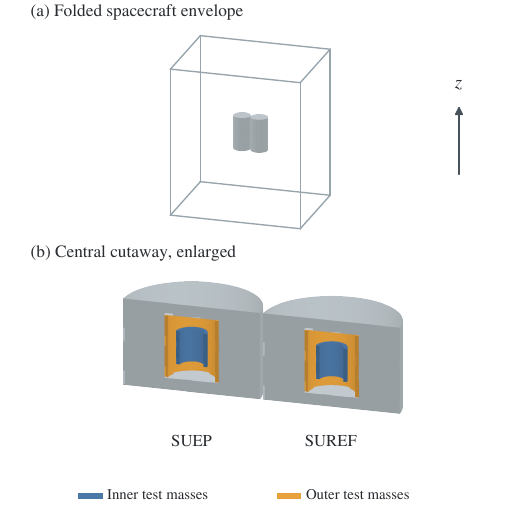}
 \caption{Simplified three-dimensional MICROSCOPE mass model.
 Panel (a) shows the folded spacecraft envelope and the two
 sensor assemblies to scale. The wireframe bounds the
 homogenized bus material outside the assembly envelopes.
 Panel (b) shows an enlarged central cutaway of SUEP and SUREF,
 with the front halves removed and the surrounding hardware
 cropped along $z$ to expose the four hollow cylindrical test
 masses. Blue and orange identify the inner and outer test
 masses, respectively, while gray represents the surrounding
 apparatus. The sensitive axis is $z$.}
 \label{fig:microscope_3d_geometry}
\end{figure}

We represent the bus by a homogeneous cuboid with the
folded spacecraft dimensions
$1.38\times1.04\times1.58\,\mathrm{m^3}$~\cite{eoPortal:MICROSCOPE}, excluding the
complete cylindrical envelopes of both sensor assemblies,
including their central cavities.
We adopt the launch mass of $301.6\,$kg, including
$16.5\,$kg of propellant~\cite{Robert:2020ddm}, as a proxy for the total
satellite mass.
Subtracting the $25\,$kg already assigned to the sensor
assemblies leaves $276.6\,$kg in the bus region.
This includes the remaining payload components, propellant,
and other spacecraft hardware.
Dividing this residual mass by the available material
volume gives a uniform density of
$120\,\mathrm{kg\,m^{-3}}$.
The three-dimensional mass distribution is illustrated
in Fig.~\ref{fig:microscope_3d_geometry}.

A more detailed model could resolve individual mass concentrations, such as the wall-mounted payload and the propellant tanks. We do not include these features because the sensor hardware is itself strongly idealized, and the available information does not allow for a consistent reconstruction of the full mass distribution. Additionally, such asymmetries can generate a local gradient even from a spatially uniform incident field. The resulting acceleration is fixed in the satellite frame, with its amplitude modulated by orbital motion rather than satellite spin. This contribution therefore appears at pure orbital harmonics, distinct from the signal considered here, and strongly depends on the detailed mass distribution. 

At leading order in the long-wavelength expansion, the
constant and linearly varying incident components propagate
according to
\begin{equation}
 \left[-\nabla^2+d\kappa^2\rho_{\rm SM}(\mathbf x)\right]
 u_a(\mathbf x)=0\,,\qquad a=0,1\,.
 \label{eq:three_dimensional_screening}
\end{equation}
Here $u_0$ and $u_1$ extend the profiles introduced in
Eq.~\eqref{eq:local_field_modes} to the three-dimensional
geometry.
For an incident component along the sensitive axis, we
impose $u_0=1$ and $u_1=z/R_{\rm exp}$ at a distant
boundary in vacuum.
Both profiles are calculated with all material regions
present, so the test masses and surrounding hardware
modify the same field.

We solve Eq.~\eqref{eq:three_dimensional_screening}
using the finite-element method.
We divide the material and vacuum regions into tetrahedral
elements and approximate each field profile by a continuous
function that is linear within each element.
To obtain the acceleration proportional to the incident
gradient of $\langle\phi^2\rangle$, we calculate the
volume-averaged derivative of the product $u_0u_1$.
For test mass $j$, the corresponding screening factor is
\begin{equation}
 T_j(d)=\frac{R_{\rm exp}}{V_j}
 \int_{V_j}d^3x\,\partial_z
 \left[u_0(\mathbf x)u_1(\mathbf x)\right],
 \label{eq:three_dimensional_test_factor}
\end{equation}
where $V_j$ is its material volume.
We evaluate this integral for all four test masses and
form $\Delta T=T_1-T_2$ separately for SUEP and SUREF.

\begin{figure}[htb]
 \includegraphics[width=\columnwidth]
 {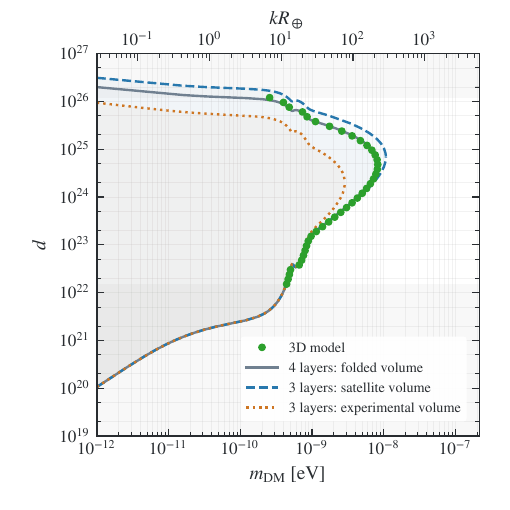}
 \caption{Joint SUEP--SUREF EP-band sensitivity for different screening
 models. Green points show the three-dimensional calculation, and the
 solid gray curve uses four concentric spherical layers. The blue dashed
 and orange dotted curves distribute the same surrounding mass uniformly
 over the folded satellite volume and the apparatus-layer volume,
 respectively.}
 \label{fig:microscope_screening_comparison}
\end{figure}

For each sampled coupling, we estimate the numerical uncertainty in $\Delta T$ by extrapolating the observed scaling under mesh refinement. We similarly use calculations with increasingly distant outer boundaries to estimate finite-domain corrections and their uncertainties. Combining mesh, finite-domain, and solver uncertainties, the estimated numerical error in $\Delta T$ is below $10\%$ for both SUEP and SUREF at every point shown in Fig.~\ref{fig:microscope_screening_comparison}.

We use a concentric spherical model to evaluate the sensitivity without repeating the three-dimensional simulation.
We represent each sensor and its surroundings by four concentric homogeneous spherical layers. 
The two inner layers represent the test masses, while
the two outer layers represent the surrounding apparatus
and the remaining satellite mass.
Keeping the apparatus and satellite bus separate retains
the concentration of matter near the sensors.
For this geometry, the field equations admit analytic solutions within each layer, with coefficients fixed by matching at the interfaces.

For the spherical geometry, the constant and linear incident components give profiles $u_0(\mathbf x)=u_0(r)$ and $u_1(\mathbf x)=u_1(r)\cos\theta$, respectively. Eq.~\eqref{eq:three_dimensional_screening} then separates into radial equations, which we solve without expanding in $d$. At leading order in $k_{\rm DM}R_{\rm exp}$, the radial profiles for $\ell=0,1$ in a layer of density $\rho_i$ obey
\begin{equation}
 \left[\frac{d^2}{dr^2}+\frac{2}{r}\frac{d}{dr}
 -\frac{\ell(\ell+1)}{r^2}-d\kappa^2\rho_i\right]u_\ell(r)=0.
 \label{eq:multilayer_radial_equation}
\end{equation}
For $d>0$, two independent solutions in a material layer
are the spherical Bessel and Hankel functions with
imaginary arguments,
\begin{equation}
 j_\ell\!\left(i\sqrt{d\kappa^2\rho_i}\,r\right),\qquad
 h_\ell\!\left(i\sqrt{d\kappa^2\rho_i}\,r\right).
 \label{eq:material_radial_basis}
\end{equation}
Here $h_\ell=j_\ell+i n_\ell$, where $n_\ell$ is the spherical
Bessel function of the second kind. In vacuum, the two
independent solutions are $r^\ell$ and $r^{-\ell-1}$.

We construct the full profiles by requiring continuity of $u_\ell$ and its radial derivative at every interface. 
In the central region, we retain only the solution regular
at the origin. 
We normalize the matched profiles
by fixing the coefficient of the constant term in $u_0$
to unity and that of the linear term in $u_1$ to
$1/R_{\rm exp}$ in the exterior vacuum region.

We insert these profiles into Eq.~\eqref{eq:shell_factor} to calculate the screening factors of the two measured test masses. Their difference, $\Delta T=T_1-T_2$, is evaluated separately for SUEP and SUREF.

We represent each cylindrical test mass by a spherical
shell with the same mass and density, using the parameters
reported in Ref.~\cite{Liorzou:2020vob}.
We take the outer radius to be that of the smallest sphere
enclosing the cylinder and determine the inner radius
from its mass.
For outer diameter $D_{{\rm out},j}$, length $L_j$,
mass $M_j$, and density $\rho_j$, this gives
\begin{equation*}
 \begin{split}
 R_{j,+}&=\frac12\sqrt{D_{{\rm out},j}^2+L_j^2},\\
 R_{j,-}&=\left(R_{j,+}^3-
 \frac{3M_j}{4\pi\rho_j}\right)^{1/3}.
 \end{split}
\end{equation*}

The apparatus and satellite bus layers have the same radii and densities for SUEP and SUREF. 
The apparatus layer carries approximately $22.5\,$kg of surrounding hardware.
This layer extends from $5.70\,$cm to $17.53\,$cm, with its outer boundary enclosing the same volume as the $360\times348\times180\,\mathrm{mm^3}$ SUEP--SUREF assembly envelope. The bus layer carries the remaining $276.6\,$kg and extends from the apparatus boundary to $81.5\,$cm. Its outer boundary encloses the same volume as the $1.38\times1.04\times1.58\,\mathrm{m^3}$ folded spacecraft envelope.
Table~\ref{tab:multilayer_parameters} lists the parameters
used in Fig.~\ref{fig:microscope_universal_bound}.

\begin{table}[htb]
\caption{Layer radii and densities of the spherical
MICROSCOPE model used in
Figs.~\ref{fig:microscope_universal_bound},
\ref{fig:microscope_screening_comparison}, and
\ref{fig:multilayer_screening_ratio}.
The apparatus and bus layers are common to both sensors.}
 \label{tab:multilayer_parameters}
 \small
 \setlength{\tabcolsep}{3pt}
 \begin{ruledtabular}
 \begin{tabular}{lcccc}
 Sensor & Layer & $R_{i,-}\,[{\rm cm}]$ & $R_{i,+}\,[{\rm cm}]$
 & $\rho_i\,[{\rm g\,cm^{-3}}]$\\
 \hline
 SUEP & 1 & 2.73 & 2.93 & 20\\
      & 2 & 5.09 & 5.29 & 4.4\\
 SUREF & 1 & 2.73 & 2.93 & 20\\
       & 2 & 5.09 & 5.29 & 20\\
 Both & 3 & 5.70 & 17.5 & 1.0\\
      & 4 & 17.5 & 81.5 & 0.12
 \end{tabular}
 \end{ruledtabular}
\end{table}

In Fig.~\ref{fig:microscope_screening_comparison}, we compare the sensitivities obtained with the spherical models and the three-dimensional calculation, using the same joint SUEP--SUREF EP-band analysis.
The four-layer model follows the broad shape of the three-dimensional sensitivity over the sampled coupling range.

The two three-layer alternatives illustrate the dependence on how the surrounding material is distributed. For example, for a homogeneous sphere of fixed mass and positive $d$, reducing the radius strengthens the suppression of the field. Both alternatives retain the two test masses explicitly inside a central cavity. The blue dashed curve distributes the surrounding mass uniformly throughout the folded satellite volume outside this cavity. The orange dotted curve concentrates the same surrounding mass within the apparatus layer representing the combined SUEP--SUREF assembly. 
The spread between these deliberately extreme models gives a qualitative indication of the uncertainty in the inferred sensitivity arising from our incomplete knowledge of the satellite mass distribution.

\begin{figure}[htb]
 \includegraphics[width=\columnwidth]
 {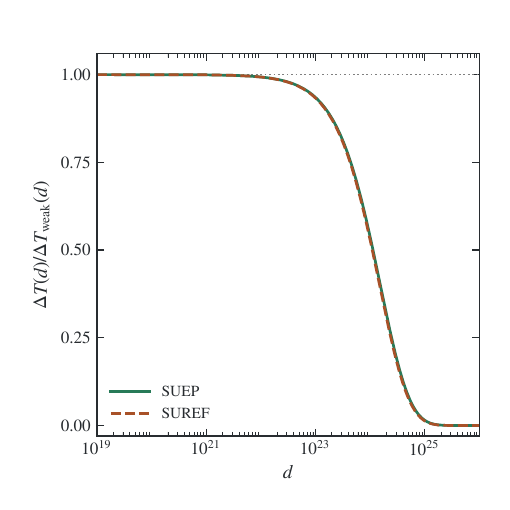}
 \caption{Ratio $\Delta T(d)/\Delta T_{\rm weak}(d)$ for the four-layer spherical MICROSCOPE model, with $\Delta T_{\rm weak}$ defined in Eq.~\eqref{eq:multilayer_weak_normalization}. The solid and dashed curves show SUEP and SUREF, respectively, while the dotted line marks unity. 
 }
 \label{fig:multilayer_screening_ratio}
\end{figure}

Fig.~\ref{fig:multilayer_screening_ratio} compares
the matched differential screening factor $\Delta T(d)$
with its weak-coupling approximation,
\begin{equation}
 \Delta T_{\rm weak}(d)
 =d\left.\frac{\partial\Delta T}{\partial d}\right|_{d=0}.
 \label{eq:multilayer_weak_normalization}
\end{equation}
We evaluate the derivative using the first-order field profiles in App.~\ref{app:weak_shells}, retaining all four layers and their finite thicknesses. At weak coupling, attenuation by the surrounding layers is negligible, and the differential signal is controlled by weak scattering from the test masses. At stronger coupling, the outer layers screen the incident field, exponentially reducing the gradient reaching the test masses in the strong-screening regime. The nearly overlapping SUEP and SUREF ratios suggest that these layers provide an approximately common suppression factor for the two sensors.

\section{MICROSCOPE signal and analysis}
\label{app:MICROSCOPE}
\setcounter{equation}{0}
\subsection{Wake force and satellite signal}

Earth scatters the ambient dark matter field, producing a
spatially varying mean-square field
$\langle\phi^2\rangle$ whose gradient generates the wake
force~\cite{Brzeminski:2026rox,Bouley:2026frx,Brzeminski:2026pgz}.
We use the profile in the strong-screening regime, where
Earth can be approximated by an opaque sphere.
After averaging over the field oscillations and halo
distribution, the profile is axisymmetric around the mean
dark matter wind direction.

Taking the mean dark matter wind direction as the polar
axis, we write the acceleration of an unscreened body as
\begin{equation}
 \mathbf a=
 \frac{d\kappa^2\rho_{\rm DM}}{2m_{\rm DM}^2R_\oplus}
 \left[
 \hat{\mathbf r}\sum_{\ell=0}^{\infty}
 b_\ell P_\ell(\cos\theta_r)
 +\hat{\boldsymbol\theta}\sum_{\ell=1}^{\infty}
 t_\ell P_\ell^1(\cos\theta_r)
 \right].
 \label{eq:appendix_wake_acceleration}
\end{equation}
Here $\theta_r$ is the angle from the mean wind direction,
and $P_\ell^1(\cos\theta_r)=
\partial_{\theta_r}P_\ell(\cos\theta_r)$.
The radial and polar coefficients are
$b_\ell=-\partial_\xi a_\ell$ and $t_\ell=-a_\ell/\xi$,
where $\xi=r/R_\oplus$ and $a_\ell$ are the Legendre
coefficients of $m_{\rm DM}^2\langle\phi^2\rangle/\rho_{\rm DM}$.

The differential acceleration in sensor $c$ is obtained
by multiplying the projection of
Eq.~\eqref{eq:appendix_wake_acceleration} onto its sensitive
axis by the screening factor $\Delta T_c$.
For a circular orbit, this gives
\begin{equation}
 \begin{aligned}
 \Delta a_c(t)={}&
 \frac{d\kappa^2\rho_{\rm DM}\Delta T_c(d)}
 {2m_{\rm DM}^2R_\oplus}\\
 &\times\sum_q A_q\cos\!\left[2\pi(f_{\rm spin}+qf_{\rm orb})t+\varphi_q\right],
 \end{aligned}
 \label{eq:appendix_harmonic_signal}
\end{equation}
where the amplitudes $A_q$ depend on the force coefficients
and the orientation of the dark matter wind relative to the
orbital plane.
The phases $\varphi_q$ are fixed by the orbital and spin
phases and the wind direction.
The $\ell=0$ force contributes at $q=1$, the EP band.
The $\ell=1$ terms populate its neighboring sidebands
$q=0,2$, while $\ell=2$ contributes at $q=-1,1,3$.

Retaining multipoles through $\ell=4$, the EP-band amplitude is
\begin{align}
 \!\!
 A_1(W)=b_0+\frac{3W^2-2}{4}b_2
 +\frac{3(35W^4-40W^2+8)}{64}b_4\,,
 \label{eq:appendix_ep_band_amplitude}
\end{align}
with the force coefficients evaluated at the satellite radius.
Here $W^2=1-(\hat{\mathbf k}\cdot\hat{\mathbf n})^2$,
where $\hat{\mathbf k}$ is the mean wind direction and
$\hat{\mathbf n}$ is the orbital normal.

\subsection{Data segments and averaging}

The orientation of the orbit of the satellite relative to the
dark matter wind changes between data segments, modifying
the predicted scalar signal in the EP band.
We therefore average the signal separately over each segment
before combining the measurements.
We use the 19 SUEP and 13 SUREF estimates of the E\"otv\"os
parameter obtained with ADAM, together with their statistical
uncertainties and systematic budgets, reported in
Ref.~\cite{Touboul:2022yrw}.

Averaging the EP-band amplitude in
Eq.~\eqref{eq:appendix_ep_band_amplitude} over segment $i$
gives
\begin{align}
 \langle A_1\rangle_i={}&b_0+
 \frac{3\langle W^2\rangle_i-2}{4}b_2\nonumber\\
 &+\frac{3}{64}
 \left(35\langle W^4\rangle_i
 -40\langle W^2\rangle_i+8\right)b_4\,.
 \label{eq:microscope_average_a1}
\end{align}

We determine the observing window for each segment from
the telemetry masks and calculate $\langle W^2\rangle_i$
and $\langle W^4\rangle_i$ by uniformly averaging the
corresponding orbital geometry.
Since $W$ varies little within each segment, these moments
satisfy $\langle W^n\rangle_i\simeq\langle W\rangle_i^n$.
Tab.~\ref{tab:microscope_session_geometry} shows the mean and root-mean-square to illustrate this stability, together with
the measurements and their statistical and systematic
uncertainties.

\begin{table*}[t]
\caption{MICROSCOPE measurements and geometric averages
for each data segment.
The ADAM estimates $\widehat\eta_{c,i}$, statistical
uncertainties $\sigma_{c,i}$, and total systematic
uncertainties $S_{c,i}$ are reported in
Ref.~\cite{Touboul:2022yrw}.
The geometric columns show the mean and root-mean-square
values of $W$ over the corresponding observing windows.}
\label{tab:microscope_session_geometry}
\begin{ruledtabular}
\begin{tabular}{llccrrr}
Sensor & Segment &
$\langle W\rangle_i$ & $\sqrt{\langle W^2\rangle_i}$ &
$10^{15}\widehat\eta_{c,i}$ &
$10^{15}\sigma_{c,i}$ &
$10^{15}S_{c,i}$ \\
\hline
SUEP & 210   & 0.83 & 0.83 & -29.2 & 13.1 & 1.8 \\
SUEP & 212   & 0.84 & 0.84 &   9.5 & 11.9 & 1.0 \\
SUEP & 218   & 0.87 & 0.87 &   6.7 &  8.1 & 1.1 \\
SUEP & 234   & 0.92 & 0.92 &   5.9 &  8.3 & 1.0 \\
SUEP & 236   & 0.94 & 0.94 &   2.6 &  6.6 & 1.2 \\
SUEP & 238   & 0.97 & 0.97 &   5.8 &  6.4 & 1.2 \\
SUEP & 252   & 0.99 & 0.99 & -14.9 &  7.3 & 1.1 \\
SUEP & 254   & 1.0  & 1.0  & -14.1 &  7.0 & 1.5 \\
SUEP & 256   & 1.0  & 1.0  &  -5.3 &  7.4 & 1.1 \\
SUEP & 326-1 & 0.86 & 0.86 & -16.3 &  9.6 & 1.6 \\
SUEP & 326-2 & 0.88 & 0.88 & -10.4 & 13.5 & 1.6 \\
SUEP & 358   & 0.95 & 0.95 &  15.8 & 10.9 & 1.1 \\
SUEP & 402   & 0.97 & 0.97 &  28.4 & 43.6 & 7.3 \\
SUEP & 404   & 0.95 & 0.95 &   4.7 &  6.7 & 1.0 \\
SUEP & 406   & 0.93 & 0.93 &   5.9 & 14.9 & 3.2 \\
SUEP & 438   & 0.83 & 0.83 & -23.4 & 24.6 & 5.5 \\
SUEP & 442   & 0.82 & 0.82 &  -1.5 & 19.1 & 7.3 \\
SUEP & 748   & 0.72 & 0.72 & -23.4 & 24.6 & 7.3 \\
SUEP & 750   & 0.72 & 0.72 &  66.9 & 38.4 & 7.3 \\
\hline
SUREF & 120-1 & 0.98 & 0.98 &  -4.2 & 15.9 & 5.4 \\
SUREF & 120-2 & 0.99 & 0.99 & -15.1 &  8.3 & 5.4 \\
SUREF & 174   & 0.82 & 0.82 &   8.0 &  4.2 & 4.0 \\
SUREF & 176   & 0.82 & 0.82 &   1.8 &  4.5 & 4.7 \\
SUREF & 294   & 0.79 & 0.79 &  -7.7 &  2.1 & 2.0 \\
SUREF & 376-1 & 1.0  & 1.0  &  -4.1 &  6.5 & 3.7 \\
SUREF & 376-2 & 1.0  & 1.0  &  -6.4 &  5.8 & 3.7 \\
SUREF & 380-1 & 0.99 & 0.99 &   7.4 &  2.4 & 0.8 \\
SUREF & 380-2 & 0.99 & 0.99 &   8.9 &  2.8 & 0.8 \\
SUREF & 452   & 0.82 & 0.82 &  -4.8 &  4.1 & 2.1 \\
SUREF & 454   & 0.82 & 0.82 &  -3.7 &  2.8 & 2.7 \\
SUREF & 778-1 & 0.86 & 0.86 &  -8.1 &  4.7 & 3.0 \\
SUREF & 778-2 & 0.87 & 0.87 &  -3.2 &  5.5 & 3.0 \\
\end{tabular}
\end{ruledtabular}
\end{table*}

\subsection{Joint recast and projected sensitivity}

We combine the SUEP and SUREF measurements with weights
set by their statistical uncertainties.
At each tested $(m_{\rm DM},d)$, the predicted segment
coefficients $\eta_{c,i}^{(1)}$ are held fixed and we fit
their common overall amplitude, giving $\widehat d$
and $\sigma_d$ in Eq.~\eqref{eq:microscope_joint_fit}.

For each measurement, $S_{c,i}$ denotes the total systematic
uncertainty reported in Tab.~10 and~11 of
Ref.~\cite{Touboul:2022yrw}.
Using the same fit weights, its contribution to the
uncertainty on the coupling is
\begin{equation}
 S_d=d\,
 \frac{\displaystyle\sum_{c,i}
 |\eta_{c,i}^{(1)}|S_{c,i}/\sigma_{c,i}^{2}}
 {\displaystyle\sum_{c,i}
 \bigl(\eta_{c,i}^{(1)}/\sigma_{c,i}\bigr)^2}\,.
 \label{eq:microscope_systematic_budget}
\end{equation}
We add the magnitudes of the individual contributions,
allowing systematic shifts with unknown signs to reinforce
one another in the fitted amplitude.

We combine $S_d$ with $\sigma_d$ in quadrature and determine
the excluded coupling range using
Eq.~\eqref{eq:microscope_budget_exclusion}.

The predicted signal grows at weak coupling but decreases
at strong coupling as the apparatus screens the incident
field. We therefore determine both the lower and upper
boundaries of the excluded coupling range.

The full-signal projection includes the harmonics
$q=-3,\ldots,5$ generated by multipoles through $\ell=4$.
It uses statistical
uncertainties only, since the reported EP-band systematic
budgets do not specify uncertainties for the additional
frequency bands.
We assume independent errors for the harmonic amplitudes,
assigning each the EP-band uncertainty of the same data
segment.
For each segment $i$, we therefore use
$\sum_q\langle A_q^2\rangle_i$ in place of
$\langle A_1\rangle_i^2$ when computing the projected
uncertainty.
Each $A_q$ includes the radial and polar contributions.

\subsection{SUEP--SUREF consistency test}

The scalar force considered here produces a differential
signal in both SUEP and SUREF, so SUREF alone does not
provide a null channel.
After accounting for their separate screening factors and
the geometry sampled by each segment, however, both sensors
should infer the same scalar signal amplitude.

At each tested $(m_{\rm DM},d)$, we restrict the sums in
Eqs.~\eqref{eq:microscope_joint_fit} and
\eqref{eq:microscope_systematic_budget} to each sensor
separately, obtaining $\widehat d_c$, $\sigma_{d,c}$,
and $S_{d,c}$.
We compare the fitted amplitudes through their difference,
\begin{align}
 \Delta\widehat d
 &=\widehat d_{\rm SUEP}-\widehat d_{\rm SUREF}\,,
 \nonumber\\
 u_{\Delta d}^2
 &=\sigma_{d,\rm SUEP}^2+\sigma_{d,\rm SUREF}^2
 \nonumber\\
 &\quad+\left(S_{d,\rm SUEP}+S_{d,\rm SUREF}\right)^2\,.
 \label{eq:microscope_sensor_contrast}
\end{align}
The independent statistical uncertainties add in quadrature,
while the systematic contributions add linearly under the
same unknown-sign prescription used above.

\begin{figure}[t]
 \includegraphics[width=\columnwidth]
 {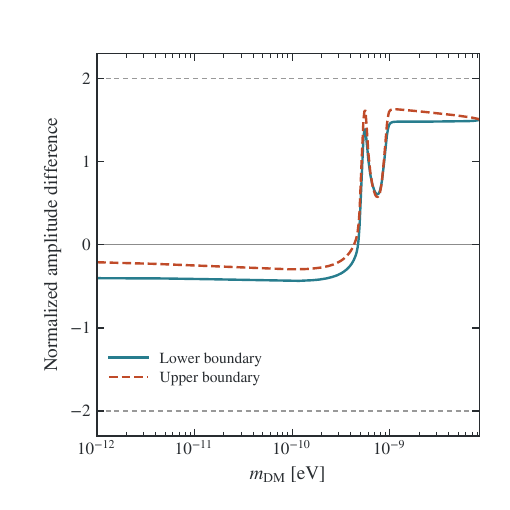}
 \caption{
 Difference between the scalar amplitudes inferred from SUEP and SUREF, normalized by their combined statistical and systematic uncertainty. The solid and dashed curves evaluate Eq.~\eqref{eq:microscope_sensor_contrast} at the lower and upper boundaries of the joint exclusion, respectively, using the four-layer model. The horizontal gray lines mark $\pm2$.
}
 \label{fig:amplitude_consistency}
\end{figure}

Fig.~\ref{fig:amplitude_consistency} shows
$\Delta\widehat d/u_{\Delta d}$ as a function of the
scalar mass.
At each mass, we evaluate the comparison at the lower
and upper boundaries of the joint exclusion.
Throughout the sampled parameters, the amplitudes inferred
from the two sensors differ by less than twice their combined
uncertainty.

\section{LAGEOS II pericenter precession}
\label{app:lageos}
\setcounter{equation}{0}
We calculate the additional pericenter precession of LAGEOS II
caused by the scalar field, using the field profile around Earth
described in Appendix~\ref{app:MICROSCOPE} and including
screening by the satellite.

For the sensitivity curves, we model LAGEOS II as two concentric homogeneous spherical regions, with outer radius $R_L=0.300\,\mathrm m$ and measured total mass
$M_L=405.38\,\mathrm{kg}$. The core has density
$8280\,\mathrm{kg\,m^{-3}}$ and a radius of approximately
$0.18\,\mathrm m$. The surrounding region represents
the aluminum housing and reflector assemblies, with its density fixed by the total satellite mass, giving
approximately $2.4\times10^3\,\mathrm{kg\,m^{-3}}$.

The spherical core has the same volume as the cylindrical brass core of diameter
$355.5\,\mathrm{mm}$ and length $232.5\,\mathrm{mm}$,
following the reconstructed satellite geometry of
Ref.~\cite{ViscoLucchesi2016}. We find that the screening factors calculated
in the spherical and cylindrical-core geometries agree to better
than $1\%$ over the parameter range tested.

We describe satellite screening by $T_L$, defined as the average
of $\nabla\langle\phi^2\rangle$ over the complete satellite,
weighted by the local mass density and normalized by the
local incident gradient.
We calculate $T_L$ using the spherical matching construction described in Appendix~\ref{app:multilayer}.
In the long wavelength limit across the satellite, $T_L$
multiplies the unscreened acceleration in
Eq.~\eqref{eq:appendix_wake_acceleration},
with $T_L=1$ without satellite screening.

The additional pericenter precession depends on the orientation
of the orbit relative to the dark matter wind.
We use the LAGEOS II orbital elements from
Ref.~\cite{Visco:2018nfj}, with semimajor axis
$s=1.216\times10^7\,\mathrm m$, eccentricity $e=0.014$,
inclination $i=52.66^\circ$, longitude of the ascending node
$\Omega_0=60.62^\circ$, and argument of pericenter
$\omega_0=251.82^\circ$.
Over the observing interval of 13 years, we approximate the
evolution of the node and pericenter by
\begin{equation}
 \Omega(t)=\Omega_0-0.63^\circ\frac{t}{\mathrm{day}},
 \qquad
 \omega(t)=\omega_0+0.44^\circ\frac{t}{\mathrm{day}}\,.
 \label{eq:lageos_orbit_evolution}
\end{equation}
The orbital elements are given in the J2000 frame at MJD~49003,
from which we measure the elapsed time $t$.

At each time $t$ and orbital angle $\vartheta$, measured from
pericenter, we evaluate the coefficients of the field around
Earth at the instantaneous orbital radius. Their angular
dependence is determined by
$\cos\theta_r=\hat{\mathbf k}(t)\cdot
\hat{\mathbf r}(t,\vartheta)$.
We use the same wind prescription as for MICROSCOPE, adding
Earth's heliocentric velocity from ephemerides to the
solar velocity. The incident direction $\hat{\mathbf k}(t)$
points opposite to the resulting velocity relative to the
dark matter halo and is expressed in the J2000 frame used
for the orbit.

For a spherically symmetric field around Earth, the scalar
force is radial, and the resulting pericenter precession can
be obtained analytically, as
in Ref.~\cite{Burrage:2026loe}.
The angular dependence of the finite wavelength profile also
produces transverse and normal components of the acceleration.
We therefore use Gauss's planetary equation to calculate the
change in the argument of pericenter.
We denote the radial and transverse acceleration components
by $a_r$ and $a_t$, and the component normal to the orbital
plane by $a_n$
\begin{align}
 \delta\dot\omega(t,\vartheta)={}&
 \frac{\tau\sqrt{1-e^2}}{2\pi se}
 \left[-a_r\cos\vartheta
 +a_t\left(1+\frac rp\right)\sin\vartheta\right]
 \nonumber\\
 &-\frac{\tau r\sin(\vartheta+\omega)\cot i}
 {2\pi s^2\sqrt{1-e^2}}\,a_n\,,
 \label{eq:lageos_gauss}
\end{align}
where $\tau=2\pi\sqrt{s^3/(G_NM_\oplus)}$ is the orbital
period, $p=s(1-e^2)$, and $r=p/(1+e\cos\vartheta)$ is the
instantaneous distance from Earth's center to the satellite.
The transverse direction lies in the orbital plane and points
toward increasing $\vartheta$, while the normal points along
the orbital angular momentum.
The final term accounts for the motion of the ascending node,
from which the argument of pericenter is measured.

We first average Eq.~\eqref{eq:lageos_gauss} over one orbital
period, holding the slowly evolving orbital orientation and
wind fixed. Weighting each orbital angle by the time the
satellite spends there gives
\begin{equation}
 \overline{\delta\dot\omega}(t)
 =\frac{1}{2\pi}\int_0^{2\pi}
 \frac{r^2}{s^2\sqrt{1-e^2}}\,
 \delta\dot\omega(t,\vartheta)\,d\vartheta\,.
 \label{eq:lageos_orbit_average}
\end{equation}
This averaged rate varies over the observing record as the
orbit precesses and the wind direction changes.

To compare with the reported precession, we calculate the
accumulated angular shift,
$\delta\omega(t)=\int_0^t
\overline{\delta\dot\omega}(t')\,dt'$.
We approximate the constant precession rate by the slope of
a straight line fitted to this angular history, allowing the
line's starting value to vary and weighting all times equally.
For an observing interval of duration $T$, this gives
\begin{equation}
 \delta\dot\omega_{\rm fit}
 =\int_0^T\frac{6t(T-t)}{T^3}\,
 \overline{\delta\dot\omega}(t)\,dt\,.
 \label{eq:lageos_slope}
\end{equation}

The orbital average can be written explicitly in terms of the
force coefficients introduced in Appendix~\ref{app:MICROSCOPE}.
The unit vector $\hat{\mathbf P}$ points toward pericenter,
$\hat{\mathbf Q}$ follows the direction of motion there, and
$\hat{\mathbf n}=\hat{\mathbf P}\times\hat{\mathbf Q}$
is the orbital normal.
The wind direction $\hat{\mathbf k}$ has components $x$, $y$,
and $z$ along these directions, respectively.
Its projection onto the orbital plane has magnitude $W$,
with $W^2=x^2+y^2=1-z^2$.

The leading terms in the eccentricity expansion, displaying
multipoles through $\ell=2$, are
\begin{align}
 \overline{\delta\dot\omega}(t)
 ={}&\frac{d\kappa^2\rho_{\rm DM}T_L\tau}
 {4\pi m_{\rm DM}^2R_\oplus s}
 \Bigg[
 b_0+\frac{\xi}{2}b_0'
 -\frac{x}{2e}(b_1+2t_1)
 \nonumber\\
 &+\frac{3W^2-2}{4}
 \left(b_2+\frac{\xi}{2}b_2'\right)
 \nonumber\\
 &+\frac{3(x^2-y^2)}{8}
 \left(3b_2+3t_2+\frac{\xi}{2}b_2'\right)
 \nonumber\\
 &-\frac32 z(x\sin\omega+y\cos\omega)\cot i\,t_2
 +\mathcal O(e)
 \Bigg].
 \label{eq:lageos_reduced_average}
\end{align}
Here $\xi=s/R_\oplus$, and primes denote derivatives with
respect to $\xi$. The coefficients and their derivatives are
evaluated at $r=s$.
We have used $b_\ell=t_\ell+\xi t_\ell'$, which follows
from their definitions.
For the sensitivity curves, we evaluate the orbital average
in Eq.~\eqref{eq:lageos_orbit_average} without expanding in
eccentricity and retain multipoles through $\ell=4$.

The radial derivatives arise because the distance from Earth
varies along the elliptical orbit.
For a monopole acceleration proportional to $1/r^2$,
the combination $b_0+\xi b_0'/2$ vanishes.
Such an acceleration changes the effective strength of gravity
without producing additional pericenter precession.
The factor $1/e$ in the dipole term reflects the fact that
the argument of pericenter becomes undefined for a circular
orbit. The acceleration itself remains finite.
The quadrupole contributes through radial and transverse
accelerations and through the motion of the ascending node.

For a fixed wind and constant field coefficients, the motion
of the ascending node and pericenter produces periodic terms
in Eq.~\eqref{eq:lageos_reduced_average}.
Their contribution to the fitted rate in
Eq.~\eqref{eq:lageos_slope} can be integrated analytically.
For a periodic term with angular frequency $\nu$ and phase
$\varphi$, we get
\begin{equation}
 \int_0^T\frac{6t(T-t)}{T^3}\,
 e^{\mathrm i(\nu t+\varphi)}\,dt
 =
 e^{\mathrm i(\varphi+\nu T/2)}
 \frac{3(\sin u-u\cos u)}{u^3},
\end{equation}
where $u=\nu T/2$.
The final factor approaches unity as $\nu\to0$.
Both the frequency and phase of each periodic term therefore
enter the fitted precession rate.

For the LAGEOS II orbit and an observing interval of 13 years,
including Earth's heliocentric velocity from ephemerids gives
\begin{align}
 \delta\dot\omega_{\rm fit}^{\rm ephem}
 \simeq{}&
 \frac{d\kappa^2\rho_{\rm DM}T_L\tau}
 {4\pi m_{\rm DM}^2R_\oplus s}
 \Big[
 1.0b_0+0.95b_0'
 +0.10b_1+0.20t_1
 \nonumber\\
 &\hspace{2.5em}
 -0.0018b_2-0.0051b_2'-0.16t_2
 \Big].
 \label{eq:lageos_ephemeris_weights}
\end{align}
These numerical factors describe the changing orientation
of the orbit relative to the wind, with the field coefficients
held fixed.

The cancellations in the plotted bound occur when the additional
precession gives zero net contribution to the fitted constant
rate over the observing interval.
In principle, sensitivity at these masses could be recovered
by analyzing the harmonics of the anomalous precession.
  We leave a dedicated analysis of this effect for future work.

\subsection{Long wavelength comparison and conditional sensitivity}

We compare the full calculation with the long wavelength
profile around an opaque Earth,
\begin{equation}
 a_0(\xi)=\left(1-\frac1\xi\right)^2,
 \qquad a_{\ell\geq1}=0.
 \label{eq:lageos_low_mass_profile}
\end{equation}
The corresponding radial acceleration produces retrograde
pericenter precession,
\begin{equation}
 \delta\dot\omega_{\rm LM}
 =-\frac{d\kappa^2\rho_{\rm DM}T_L\tau R_\oplus^2}
 {4\pi m_{\rm DM}^2s^4(1-e^2)}.
 \label{eq:lageos_low_mass_precession}
\end{equation}
We also extrapolate this expression to higher masses, beyond
the long wavelength regime, to compare its predicted
sensitivity with the full calculation.

We compare the predicted rate $\delta\dot\omega_{\rm fit}$
with the anomalous rate
$\Delta_{\rm obs}=-0.40\,\mathrm{mas\,yr^{-1}}$ quoted in
Ref.~\cite{Burrage:2026loe}, based on the analysis of
Lucchesi and Peron~\cite{Lucchesi:2014uza}.
We add the quoted statistical and systematic errors linearly
and use twice their sum to define the approximate sensitivity
boundary,
\begin{equation}
 \left|\delta\dot\omega_{\rm fit}-\Delta_{\rm obs}\right|
 =181\,\mathrm{mas\,yr^{-1}}.
 \label{eq:lageos_threshold}
\end{equation}

\begin{figure}[t]
 \centering
 \includegraphics[width=\columnwidth]{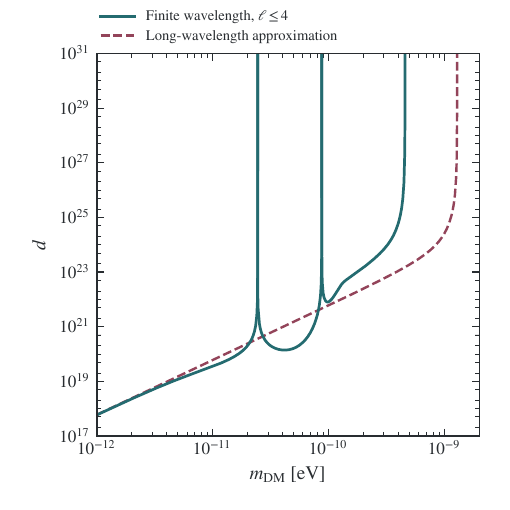}
\caption{LAGEOS II sensitivity from the long wavelength
expression extrapolated to higher masses, shown by the dashed burgundy curve, and the
finite wavelength field profile around Earth through $\ell=4$, shown by the solid teal curve.
Both curves include satellite screening and use the sensitivity
criterion in Eq.~\eqref{eq:lageos_threshold}.
Narrow gaps correspond to cancellations in the fitted constant
precession rate.}
 \label{fig:lageos_comparison}
\end{figure}

Fig.~\ref{fig:lageos_comparison} shows that the calculation
at finite wavelength, retaining multipoles through $\ell=4$,
reaches $m_{\rm DM}\simeq4.6\times10^{-10}\,\mathrm{eV}$,
compared with $1.3\times10^{-9}\,\mathrm{eV}$ from the
extrapolated long wavelength expression.
Therefore, including the wavelength dependence reduces the mass reach
by a factor of about $3$.

\bibliographystyle{apsrev4-2}
\bibliography{references}

\end{document}